# Theoretical Study on the Structural and Thermodynamic Properties of U-He compounds under High Pressure†

Ye Cao,[a] Hongxing Song,*[a] Xiaozhen Yan,[a] Hao Wang,[a] Yufeng Wang,[a] Fengchao Wu,[a] Leilei Zhang,[a] Qiang Wu*[a] and Huayun Geng*[ab]

Uranium is considered as a very important nuclear energy material because of the huge amount of energy released. As the main products of spontaneous decay of uranium, helium is difficult to react with uranium for its chemical inertness. Therefore, bubbles will be formed inside uranium, which could greatly reduce the performance of uranium or cause the safety problems. Additionally, nuclear materials are usually operated in an environment of high-temperature and high-pressure, so it is necessary to figure out the exact state of helium inside uranium at extreme conditions. Here, we explored the structural stability of U-He system under high-pressure and high-temperature by using density functional theory calculations. Two metastable phases are found between 50 and 400 GPa: U$_4$He with space group $Fmmm$ and U$_6$He with space group $P\bar{1}$. Both are metallic and adopt layered structures. Electron localization function calculation combined with charge density difference analysis indicate that there are covalent bonds between U and U atoms in both $Fmmm$-U$_4$He and $P\bar{1}$-U$_6$He. Compared with the elastic modulus of α-U, the addition of helium has certain influence on the mechanical properties of uranium. Besides, first-principles molecular dynamics simulations were carried out to study the dynamical behavior of $Fmmm$-U$_4$He and $P\bar{1}$-U$_6$He at high-temperature. It is found that $Fmmm$-U$_4$He and $P\bar{1}$-U$_6$He undergo one-dimensional superionic phase transitions at 150 GPa. Our study revealed exotic structure of U-He compounds beyond the form of bubble under high-pressure and high-temperature, that might be relevant to the performance and safety issue of nuclear materials at extreme conditions.

## Introduction

The search for environment-friend energy storage materials is essential for sustainable development.[1,2] Carbon-based energy is the most widely used in modern society. However, carbon dioxide, as the main combustion products when carbon-based energy is released, will accumulate in the air, result the greenhouse effect, and produce unexpected weather in the future.[3-5] Nuclear energy is much cleaner and more efficient than fossil fuels. The energy density of nuclear energy is millions of times higher than that of fossil fuel and it does not emit huge amounts of pollutants as well. The application of nuclear energy plays a key role in low-carbon emission reduction and is an important pillar to solve the problem of future energy supply.[6,7] As a typical actinide metal, uranium has been widely used in the field of nuclear energy and has aroused strong research enthusiasm for its unique nuclear properties.[8-11] The actinide metals voluntarily decay to inert gases because of their radioactivity, e.g., $^{238}$U decays into helium: $^{238}_{92}$U$\rightarrow$$^{234}_{90}$Th $+$ $^4_2$He. Due to its chemical inactivity, helium is insoluble in uranium.

Therefore, helium could easily diffuse and accumulate inside the bulk of uranium to form helium bubbles, which causes the swelling of materials.[12,13] The existence of helium bubbles at ambient conditions will lead to a change in the mechanical properties of nuclear materials, which will seriously affect the safety of nuclear energy.

Uranium has much broad applications than as the nuclear fuel in engineering industry, some of them will experience high-pressure and high-temperature. Pressure can reduce the distance between atoms, change their bonding nature, and adjust the relative stability of different structure. Thus the structural, electronic and chemical properties of materials under high-pressure could be quite different from the pictures in ambient conditions.[14-18] For example, helium behaves as a typical inert element at ambient pressure, however, scientists have been surprised to find stable helium compounds under high pressure in recent years, such as LiHe,[19] Na$_2$He,[20] FeHe,[21] FeO$_2$He,[22] MgOHe,[23] He$_2$H$_2$O,[24] NH$_3$He.[25] These results proved that pressure can change the chemistry of helium to form unprecedented and stable compounds. It is interesting to ask whether helium can exist in uranium in another form at high pressure, which is different from the usually observed bubble. If true, the currently hypothesized behavior of nuclear materials at extreme conditions might be modified dramatically.

In this work, we investigated the possible polymorphism of U-He system, explored their structural stability and thermodynamic properties under high pressure and


[a] National Key Laboratory of Shock Wave and Detonation Physics, Institute of Fluid Physics, China Academy of Engineering Physics, Mianyang, Sichuan 621900, P. R. China.

[b] HEDPS, Center for Applied Physics and Technology, and College of Engineering, Peking University, Beijing 100871, P. R. China.

† Electronic Supplementary Information (ESI) available.



temperature. Through crystal structural prediction, the structures of $U_nHe$ (n=1, 2, 3, 4, 6) up to 150 GPa were predicted based on density functional theory calculations. Two metastable structures ($U_4He$ with space group *Fmmm* and $U_6He$ with space group $P\bar{1}$) were identified, and their bonding properties have been calculated and analyzed, including electron localization function (ELF) and Bader charge calculation. In order to understand the effect of the addition of helium on the mechanical properties of uranium, the elastic modulus of two structures were also calculated. Moreover, the dynamical behaviors of *Fmmm*-$U_4He$ and $P\bar{1}$-$U_6He$ at high-temperature were analyzed. In these studies, we proved that uranium and helium could possibly form metastable structures, in addition to the bubble form under high pressure. We referred metastable structures as those thermodynamically metastable (higher in enthalpy or free energy), but dynamically stable (without any imaginary frequencies in phonon dispersions). The metastable structures are usually formed in non-equilibrium conditions which can be achieved by shock compression and rapid quenching, even if the structures do not have a stable phase region in the equilibrium phase diagram, they could be generated by high-pressure and/or high-temperature experiments and observed due to their metastability.

Nowadays, many experiments have proved that metastable structures can be observed in real world.[26-32] Hence, our work may provide guidance for the utilization of uranium-based nuclear energy.

## Computational details

The ground state structures of $U_nHe$ (n=1, 2, 3, 4, 6) at 150 GPa were predicted by particle swarm optimization (PSO) method as implemented in CALYPSO program,[33] with a maximum number of 4 formula units (f.u.) in the unit cell. For each ratio of the compositions, 900 structures were generated, the total number of generated structures of the U-He system is ~4500. The structure searches are considered to be converged if there is no new lowest-enthalpy structure after generating ~600 structures for each ratio of compositions. Total energy calculations were based on density functional theory[34] by using the VASP code[35,36] within the project augmented wave (PAW) method.[37] The generalized gradient approximation (GGA) in the form of the Perdew-Burke-Ernzerhof (PBE) functional was adopted for the exchange-correlation functional.[38] The plane wave basis cut-off energy was set as 650 eV. The valence electron configuration of

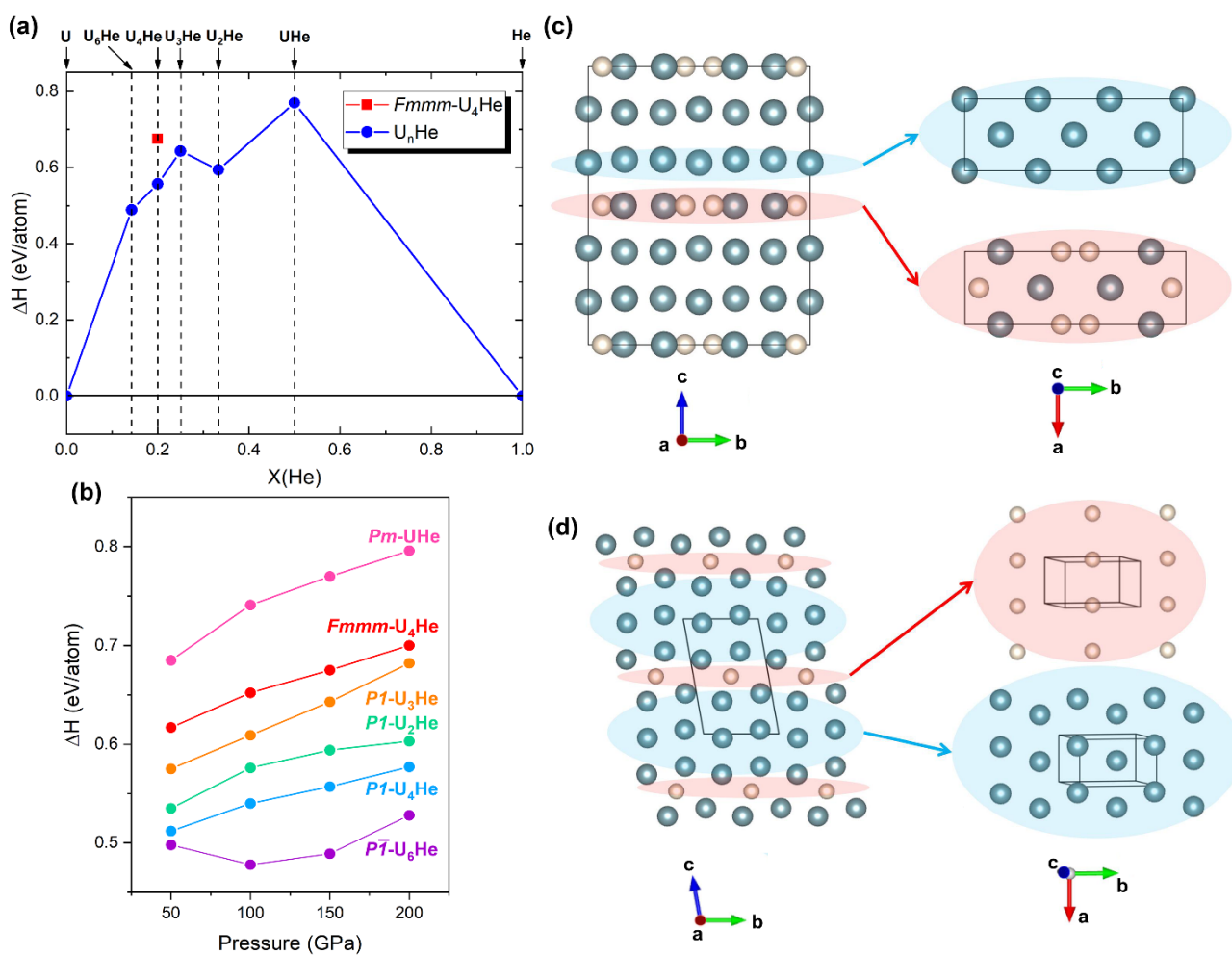

**Fig. 1.** (a) Calculated formation enthalpy ΔH of $U_nHe$ compounds at 150 GPa; (b) Pressure-dependence of the formation enthalpies ΔH of $U_nHe$ between 50 and 200 GPa, and crystal structure of (c) *Fmmm*-$U_4He$ and (d) $P\bar{1}$-$U_6He$ at 150 GPa.



U and He atoms is $6s^2 6p^6 5f^3 6d^1 7s^2$ and $1s^2$, respectively. The Brillouin zone is sampled by the Γ-centered mesh with a k-point spacing of $2\pi \times 0.03$ Å$^{-1}$. For the reference enthalpy of pure uranium and helium, we used $\alpha$-U and $hcp$-He. The phonon dispersion spectrum was calculated by using PHONOPY program.[39] The LO-TO splitting is quite small in such materials, which could be neglected here. The isothermal isobaric ensemble ($NPT$) was used in the first-principles molecular dynamics (FPMD) simulations of $Fmmm$-U$_4$He and $P\bar{1}$-U$_6$He under finite temperature. The time step was set as 1 fs, and the total simulation time was 10 ps for each simulation (except for the total simulation time of 100 ps for 300~900 K). In the supercell of the FPMD simulation box, there are 96 (24) U (He) atoms for $Fmmm$-U$_4$He, and 96 (16) U (He) atoms for $P\bar{1}$-U$_6$He.

## Results and discussion

In order to understand the physical and chemical behavior of U-He system at high temperature and pressure, we first explored their high-pressure structural stability at zero temperature. We selected 150 GPa as the chosen pressure for structure prediction by referring to the stable pressure range of Na$_2$He (100~200 GPa).[20] At 150 GPa, the lowest-energy structures with different stoichiometry have been predicted, including UHe with space group $Pm$ ($Pm$-UHe), U$_2$He with space group $P1$ ($P1$-U$_2$He), U$_3$He with space group $P1$ ($P1$-U$_3$He), U$_4$He with space group $P1$ ($P1$-U$_4$He). Fig. 1(a) shows the calculated formation enthalpy of U$_n$He. The formation enthalpy ΔH is defined as:

$$\Delta H = \frac{H(\mathrm{U}_n\mathrm{He}) - nH(\mathrm{U}) - H(\mathrm{He})}{n+1} \qquad (1)$$

where $H$(U) and $H$(He) are the enthalpy of the most stable structure of element uranium ($\alpha$-U) and helium ($hcp$-He), respectively. From Fig. 1(a), all these ordered structures have positive formation enthalpy within 200 GPa, suggesting that U and He will segregate and form He bubble. However, besides this morphology, we find two metastable phases that can be existed at high-pressure, one of them is U$_4$He with space group $Fmmm$ ($Fmmm$-U$_4$He), which has a formation enthalpy of 0.675 eV/atom, another one is U$_6$He with space group $P\bar{1}$ ($P\bar{1}$-U$_6$He). Their metastable pressure ranges are described in detail later.

It is interesting to note that there is He$_2$ dimer in $Fmmm$-U$_4$He, which resembles the O$_2$ dimer existence as interstitial defects in UO$_2$.[40] In fact, this structure can be viewed as a distorted face-centered cubic ($fcc$) phase of uranium in which some uranium atoms are replaced by He$_2$ dimer, and the shortest He-He distance is 1.3 Å. That is, it is an ordered substitutional compound of U-He$_2$. By contrast, there is no He dimer in $P\bar{1}$-U$_6$He. In this latter phase, the atomic He inserted into the uranium matrix, forming atomic He layer sandwiched by U atoms. The lattice parameters and atomic sites of $Fmmm$-U$_4$He and $P\bar{1}$-U$_6$He are summarized in Table S1. It is worthwhile noting that neither the structure of $Fmmm$-U$_4$He and $P\bar{1}$-U$_6$He, nor their uranium matrix bear any resemblance with the uranium-based alloys or compounds, as well as the high-pressure phase of pure uranium,[41, 42] indicating they are new structure unknown by far.

Phonon dispersion calculation shows that there are imaginary frequencies in $Pm$-UHe, $P1$-U$_2$He, $P1$-U$_3$He, and $P1$-U$_4$He, proving that these structures are not dynamically stable between 100 to 300 GPa (see Fig. S1 in SI). However, as shown

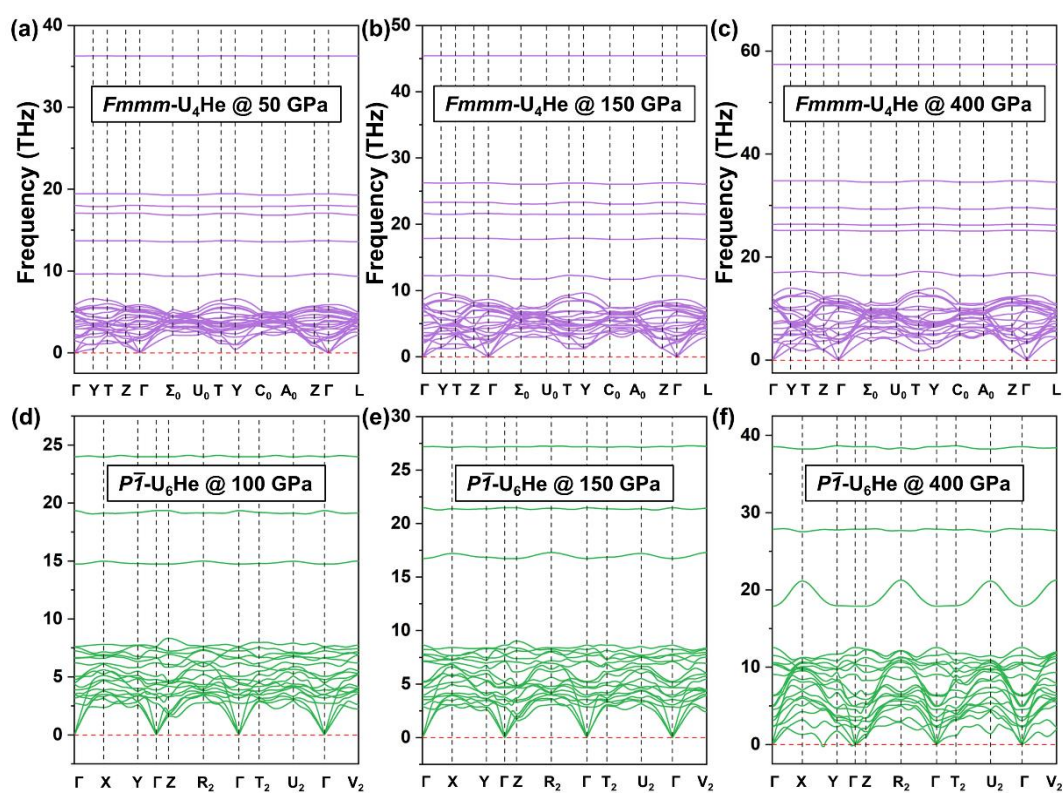

**Fig. 2.** The phonon dispersion spectra at selected pressures of $Fmmm$-U$_4$He and $P\bar{1}$-U$_6$He.



in Fig. 2 (a)-(c), *Fmmm*-U$_4$He is dynamically stable above 50 GPa, and no imaginary modes from 50 up to 400 GPa. By contrast, there is no imaginary frequency in *P$\overline{1}$*-U$_6$He, and it is dynamically stable from 100 GPa to 400 GPa (see Fig. 2(d)-(f)). We also examined their thermodynamical stability at finite temperatures within the quasi-harmonic approximation (QHA).[43] As illustrated in Fig. S2, although the calculated Gibbs free energy of *Fmmm*-U$_4$He and *P$\overline{1}$*-U$_6$He is still positive up to 1000 K, it shows a decreasing trend and the formation Gibbs free energy reduces by a magnitude of 0.02 eV/atom approximately within 1000 K.

The calculated electronic band structure and projected density of state (PDOS) of *Fmmm*-U$_4$He and *P$\overline{1}$*-U$_6$He at 150 GPa are shown in Fig. 3. The absence of an energy gap at the Fermi level indicates that both *Fmmm*-U$_4$He and *P$\overline{1}$*-U$_6$He are metallic at 150 GPa. Their PDOS results reveal that near the Fermi level, the main contribution is from U-d and U-f orbitals. The electron localization function (ELF) is one of the methods to characterize the localization degree of electrons. The value ELF=1 represents

highly localization of electrons, the value of ELF close to 0 corresponds to the region of the space where there is a low probability of finding electron localization, the value ELF=0.5 corresponds to a region of electron gas-like behaviour.[44] We combined ELF calculation, charge density difference and Bader charge analysis (see Table S2) to identify the bonding nature of these two structures. As depicted in Fig. 3(c) and (g), the value of ELF is about 0.7 around U atoms, and 1 around He atoms, which means electrons are highly localized around He atoms in both *Fmmm*-U$_4$He and *P$\overline{1}$*-U$_6$He. Furthermore, according to the calculated Bader charge, each He atom obtains only 0.1 e charge from U atoms, which is too small to define a prominent ionic bond between U and He atoms. However, from the calculated ELF and charge density difference between U and U atoms, it can be seen that there are covalent bonds between neighboring U atoms in both *Fmmm*-U$_4$He and *P$\overline{1}$*-U$_6$He, very similar to the situation observed in uranium-based transition metal alloys.[41,42] We also note the strong anti-bonding feature of the He$_2$ dimer in *Fmmm*-U$_4$He. It is in sharp contrast the

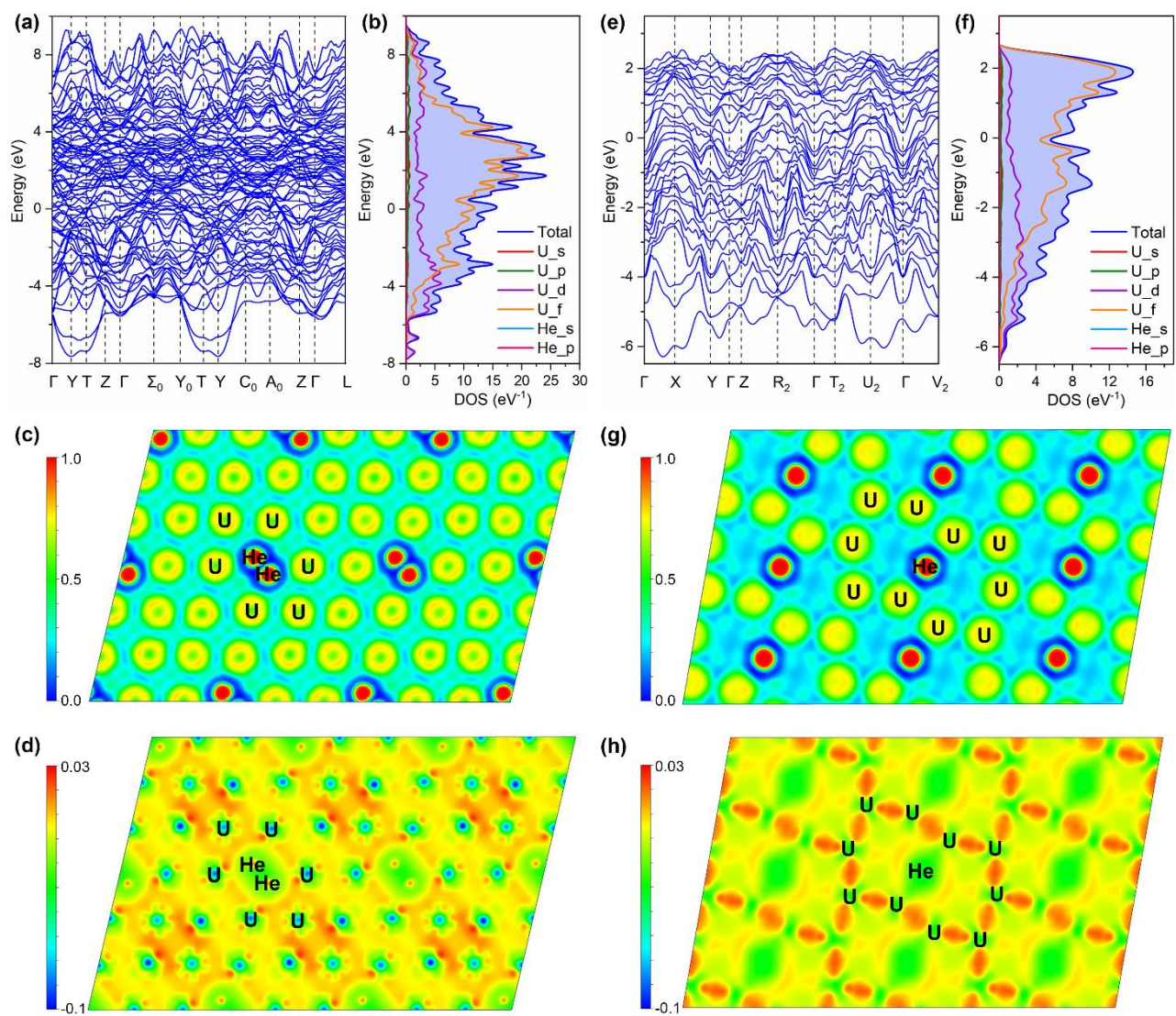

**Fig. 3.** (a) and (e) Calculated electronic band structure, (b) and (f) projected density of state (PDOS), (c) and (g) the electron localization function (ELF), and (d) and (h) charge density difference of *Fmmm*-U$_4$He (left panels) and *P$\overline{1}$*-U$_6$He (right panels) at 150 GPa, respectively.



**Table 1.** The calculated independent elastic constants $C_{ij}$, bulk modulus $B$, shear modulus $G$, Young's modulus $E$ (in GPa), Poisson's ratio $v$, longitudinal $v_l$, transverse $v_t$, average sound velocities $v_m$ (in m/s), Debye temperature $\theta_D$ (in K), and $B/G$ of $Fmmm$-U$_4$He and $P\bar{1}$-U$_6$He at 100 GPa.

| | | | | | $Fmmm$-U$_4$He | | | | |
|---|---|---|---|---|---|---|---|---|---|
| Pressure (GPa) | $C_{11}$ | $C_{22}$ | $C_{33}$ | $C_{44}$ | $C_{55}$ | $C_{66}$ | $C_{12}$ | $C_{13}$ | $C_{23}$ |
| 100 | 1003.9 | 1002.1 | 742.6 | 179.8 | 144.0 | 359.0 | 313.7 | 259.8 | 283.1 |
| Pressure (GPa) | $B$ | $G$ | $E$ | $v$ | $v_l$ | $v_t$ | $v_m$ | $\theta_D$ | $B/G$ |
| 100 | 489.8 | 246.1 | 632.4 | 0.285 | 5669.3 | 3109.9 | 3467.1 | 445.0 | 1.990 |

| | | | | | $P\bar{1}$-U$_6$He | | | | |
|---|---|---|---|---|---|---|---|---|---|
| Pressure (GPa) | $C_{11}$ | $C_{22}$ | $C_{33}$ | $C_{44}$ | $C_{55}$ | $C_{66}$ | $C_{12}$ | $C_{13}$ | $C_{14}$ |
| | 894.5 | 812.7 | 967.9 | 326.5 | 287.2 | 268.3 | 304.2 | 349.1 | -10.4 |
| | $C_{15}$ | $C_{16}$ | $C_{23}$ | $C_{24}$ | $C_{25}$ | $C_{26}$ | $C_{34}$ | $C_{35}$ | $C_{36}$ |
| 100 | -67.9 | -21.1 | 292.8 | -17.8 | 18.1 | -110.2 | 65.4 | 101.0 | 62.3 |
| | $C_{45}$ | $C_{46}$ | $C_{56}$ | | | | | | |
| | 27.1 | 52.9 | 7.8 | | | | | | |
| Pressure (GPa) | $B$ | $G$ | $E$ | $v$ | $v_l$ | $v_t$ | $v_m$ | $\theta_D$ | $B/G$ |
| 100 | 502.6 | 277.5 | 703.2 | 0.267 | 5785.3 | 3262.5 | 3629.2 | 459.2 | 1.811 |

obvious bonding of O$_2$ dimer in UO$_2$[40]. This phenomenon indicates that the He$_2$ dimer is not stabilized by chemical bonding of itself, but rather stabilized by the compression effects of surrounding uranium matrix.

To further explore the influence of helium's participation on the mechanical characteristics of uranium, the elastic constants of $Fmmm$-U$_4$He and $P\bar{1}$-U$_6$He were calculated by the energy-strain method as implemented in VASPKITand MyElas.[45,46] Table S3 summarizes our calculated elastic constants and elastic modulus of $\alpha$-U under 0 and 100 GPa, which are in good agreement with the data from experimental[47] and other theoretical calculations.[48-50] Then we calculated the independent elastic constants, bulk modulus, shear modulus, Young's modulus, Poisson's ratio, and longitudinal, transverse, average sound velocities, and Debye temperature $\theta_D$ of $Fmmm$-U$_4$He and $P\bar{1}$-U$_6$He at 100 GPa, respectively (as listed in Table 1). On the basis of Born stability criteria,[51] such as the requirements for orthogonal structures:

$$C_{11} > 0, C_{44} > 0, C_{55} > 0, C_{66} > 0, C_{11}C_{22} > C_{12}^2,$$

$$C_{11}C_{22}C_{33} + 2C_{12}C_{13}C_{23} - C_{11}C_{23}^2 - C_{22}C_{13}^2 - C_{33}C_{12}^2 > 0 \quad (2)$$

the calculated $C_{ij}$ of $Fmmm$-U$_4$He (and $P\bar{1}$-U$_6$He) satisfy the stability criteria of orthogonal (triclinic) structures, suggesting that both $Fmmm$-U$_4$He and $P\bar{1}$-U$_6$He are mechanical stable at 100 GPa.

Young's modulus is an index to measure the difficulty of elastic deformation of materials, and can be reflect the stiffness of materials. At 100 GPa, the Young's modulus of $Fmmm$-U$_4$He ($P\bar{1}$-U$_6$He) is 632.4 (703.2) GPa, which is smaller than that of $\alpha$-U (822.4 GPa), indicating that the addition of He reduces the hardness of U, and the hardness decreases with the increase of He content. Besides, the increase of He content reduces the incompressibility and plastic deformation resistance of uranium, which is reflected in the reduced bulk modulus $B$ and shear modulus $G$ of $Fmmm$-U$_4$He and $P\bar{1}$-U$_6$He. According to Pugh criterion,[52] the larger the value of $B/G$, the higher the ductility of materials. At 100 GPa, $B/G$ ($\alpha$-U) =1.547, which means that $\alpha$-U is brittle. However, the value of $B/G$ for $Fmmm$-U$_4$He ($P\bar{1}$-U$_6$He) is 1.990 (1.811), illustrated that both $Fmmm$-U$_4$He and $P\bar{1}$-U$_6$He behaves ductile. The Poisson's ratio of $Fmmm$-U$_4$He and $P\bar{1}$-U$_6$He is 0.285 and 0.267, higher than that of $\alpha$-U (0.234), proving that the addition of He improves the ductility of U.

Above analysis suggests that the helium atoms are weakly coupling to the uranium matrix, implying the possibility of forming superionic state of helium at high-temperature. To verity this hypothesis, we investigated the dynamical properties of $Fmmm$-U$_4$He and $P\bar{1}$-U$_6$He at 150 GPa by the FPMD simulations. The calculated mean squared displacement (MSD) and the trajectories of $Fmmm$-U$_4$He and $P\bar{1}$-U$_6$He are depicted in Fig. 4. For $Fmmm$-U$_4$He (as shown in Fig. 4(a), (b), (e), (f)) at room temperature, it remains in a solid phase as evidenced by the null diffusion coefficients of U and He atoms ($D_U = D_{He} = 0$).



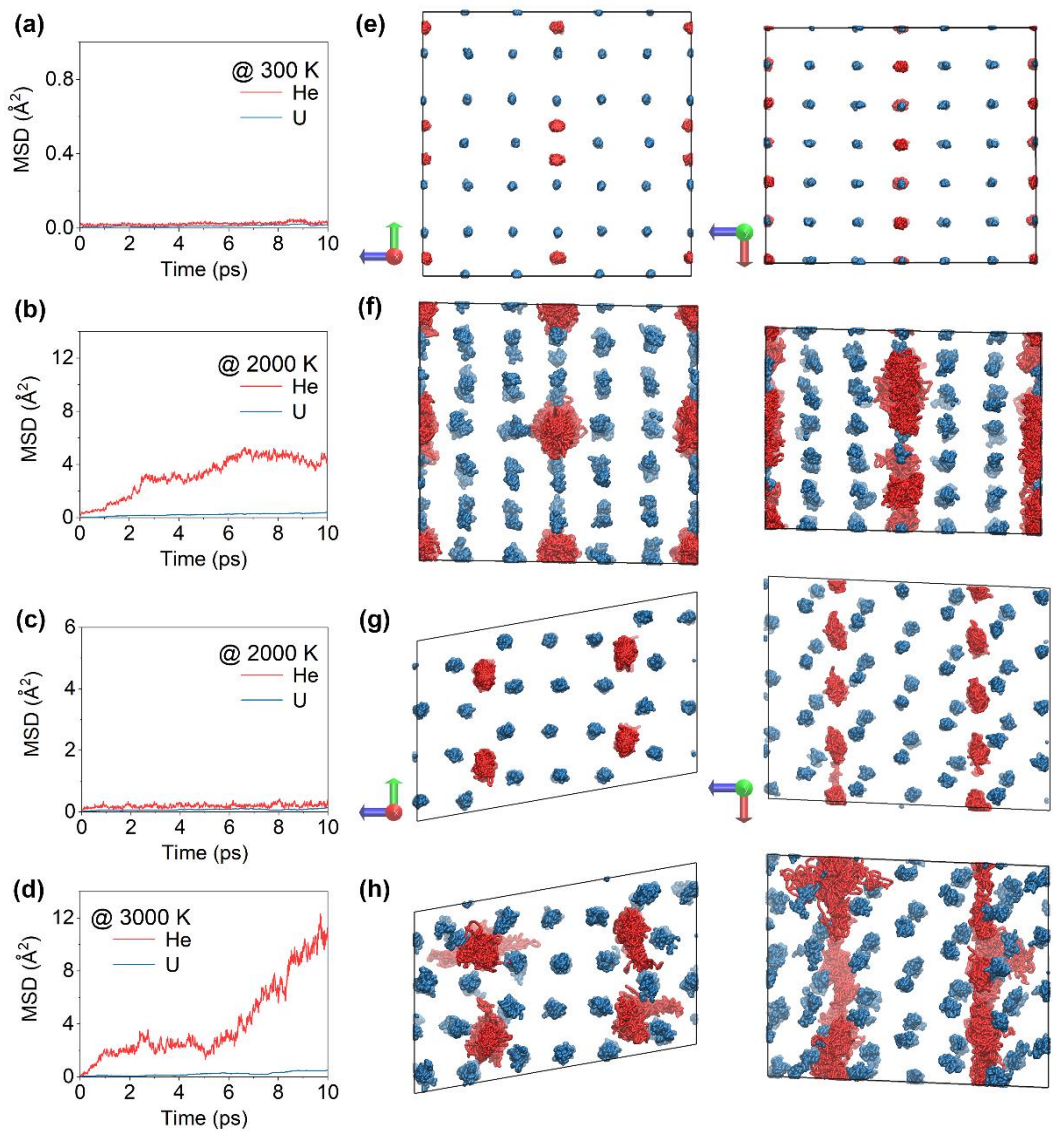

**Fig. 4.** Dynamical behavior of U and He atoms in (a, b, e, f) *Fmmm*-U₄He and (c, d, g, h) *P*1̄-U₆He at 150 GPa. (a)-(d) The calculated mean squared displacement (MSD) and (e)-(h) the trajectories of He (red) and U (blue) atoms from the FPMD simulations, respectively.

The atomic trajectories show that all atoms vibrate around the original sublattice. When temperature rises to 2000 K, U atoms still vibrate around their equilibration position; however, He atoms start diffuse significantly along the [110] direction with a non-zero diffusion coefficient ($D_U = 0$, $D_{He} > 0$). This means that *Fmmm*-U₄He begins to exhibit a one-dimensional superionic (named as 1DSI) state at the temperature of 2000 K. As the temperature continues to rise to 4000 K, as displayed in Fig. S3 of SI, both U and He atoms diffuse freely as in liquids. The phase transition sequence (solid-superionic-liquid) of *P*1̄-U₆He is similar to that of *Fmmm*-U₄He (see Fig. 4(c), (d), (g), (h)), it behaves as a typical solid phase at 2000 K, then transforms into the 1DSI phase along the [001] direction at around 3000 K, and completely melts above 5000 K as shown in Fig. S4 of SI. It is evident from the crystalline structure that the He atoms are not shielded by U atoms in the direction of [110] (for *Fmmm*-U₄He)

and [001] (for *P*1̄-U₆He), respectively, resulting in low potential barrier and easier diffusion. For this physical reason, the 1DSI phenomenon of the two structures occurs. To further discuss the stability of *Fmmm*-U₄He and *P*1̄-U₆He below the melting point, the FPMD simulations in a longer time (100 ps) were performed. The trajectories of *Fmmm*-U₄He and *P*1̄-U₆He show that all atoms vibrate around the original sublattice, indicating that both of the two structures remain the solid phase and no phase transition occurs at 300, 500, 700, and 900 K (see Fig. S5). The evolution of enthalpy with time illustrates that the system of *Fmmm*-U₄He and *P*1̄-U₆He reach the equilibrium state very quickly, and then oscillates around the equilibrium value, as depicted in Fig. S6. The radial distribution function (RDF) of *Fmmm*-U₄He and *P*1̄-U₆He confirms that there is no solid-solid phase transition occurred in *Fmmm*-U₄He and *P*1̄-U₆He between 300 and 900 K (see Fig. S7). These results of FPMD





simulations prove the structural stability of both $Fmmm$-$U_4$He and $P\bar{1}$-$U_6$He.

## Conclusions

In summary, we obtained the ground state and metastable structures of $U_n$He (n=1, 2, 3, 4, 6) at 150 GPa, by combining with crystal structural prediction and density functional theory. The calculated phonon dispersions reveal that $Fmmm$-$U_4$He and $P\bar{1}$-$U_6$He are dynamically stable, showing ordered polymorphs other than the usually spherical helium bubble or layered (planar clustering) configuration[53] are possible under extreme conditions. Both $Fmmm$-$U_4$He and $P\bar{1}$-$U_6$He exhibit layered structures and are metallic in electronic structure. The ELF and charge density difference analysis confirm that there are covalent bonds between neighboring U atoms. The calculated elastic modulus results demonstrate that the participation of He has noticeable effect on the mechanical properties of $\alpha$-U. Furthermore, both $Fmmm$-$U_4$He and $P\bar{1}$-$U_6$He exhibit one-dimensional superionic phase transition. These findings reveal the possibility of forming metastable U-He compounds under high-pressure, which drastically modifies the properties of uranium. This not only provides us a deeper understanding of the nuclear energy materials but also will motivate subsequent experimental research.

## Author Contributions


Ye Cao,[a] and Hua Y. Geng[*ab] designed and performed theoretical calculations, and analyzed data. Hong X. Song,[*a] Xiao Z. Yan,[a] Hao Wang,[a] Yu F. Wang,[a] Feng C. Wu,[a] Lei L. Zhang,[a] assisted in theoretical calculations. Hong X. Song,[*a] Qiang Wu[*a] and Hua Y. Geng[*ab] provided intellectual input. Ye Cao,[a] Hong X. Song,[*a] and Hua Y. Geng[*ab] wrote the manuscript.


## Conflicts of interest

The authors declare no competing financial interests.

## Acknowledgements


This work is supported by National Key R&D Program of China under Grant No. 2021YFB3802300, the National Natural Science Foundation of China under Grant No. 12202418 and the NSAF under Grant No. U1730248. Part of the computation was performed using the supercomputer at the Center for Computational Materials Science (CCMS) of the Institute for Materials Research (IMR) at Tohoku University, Japan.

Supporting Information

*for*

# Theoretical Study on the Structural and Thermodynamic Properties of U-He compounds under High Pressure


Ye Cao,[a] Hongxing Song,*[a] Xiaozhen Yan,[a] Hao Wang,[a] Yufeng Wang,[a] Fengchao Wu,[a] Leilei Zhang,[a] Qiang Wu*[a] and Huayun Geng*[ab]

a National Key Laboratory of Shock Wave and Detonation Physics, Institute of Fluid Physics, China Academy of Engineering Physics, Mianyang, Sichuan 621900, P. R. China;

b HEDPS, Center for Applied Physics and Technology, and College of Engineering, Peking University, Beijing 100871, P. R. China

**Corresponding Authors:**

hxsong555@163.com (Hong X. Song);

wuqianglsd@163.com (Qiang Wu);

hygeng@outlook.com (Hua Y. Geng)




**Table S1.** Crystal structure information of *Fmmm*-U$_4$He and *P$\bar{1}$*-U$_6$He at 150 GPa.

| Phase | Space group | Lattice parameters | Atom | Fractional coordinates | | | Wyckoff Position |
|-------|-------|-------|-------|-------|-------|-------|-------|
| | | | | x | y | z | |
| U$_4$He | *Fmmm* | a=10.320 Å | U1 | 0.664 | 0.500 | -0.165 | 16n |
| | | b=3.434 Å | U5 | 0.341 | 0.500 | -0.500 | 8g |
| | | c=13.017 Å | U7 | 0.500 | 0.500 | -0.346 | 8i |
| | | α=β=γ=90° | He1 | 0.936 | 0.500 | -0.500 | 8g |
| U$_6$He | *P$\bar{1}$* | a=2.746 Å | | | | | |
| | | b=4.481 Å | U1 | 0.533 | 0.676 | 0.716 | 2i |
| | | c=7.023 Å | U2 | 0.801 | 0.853 | 0.353 | 2i |
| | | α=100.0° | U4 | 0.819 | 0.241 | 0.964 | 2i |
| | | β=92.3° | He1 | 0.000 | 0.500 | 0.500 | 1g |
| | | γ=90.7° | | | | | |



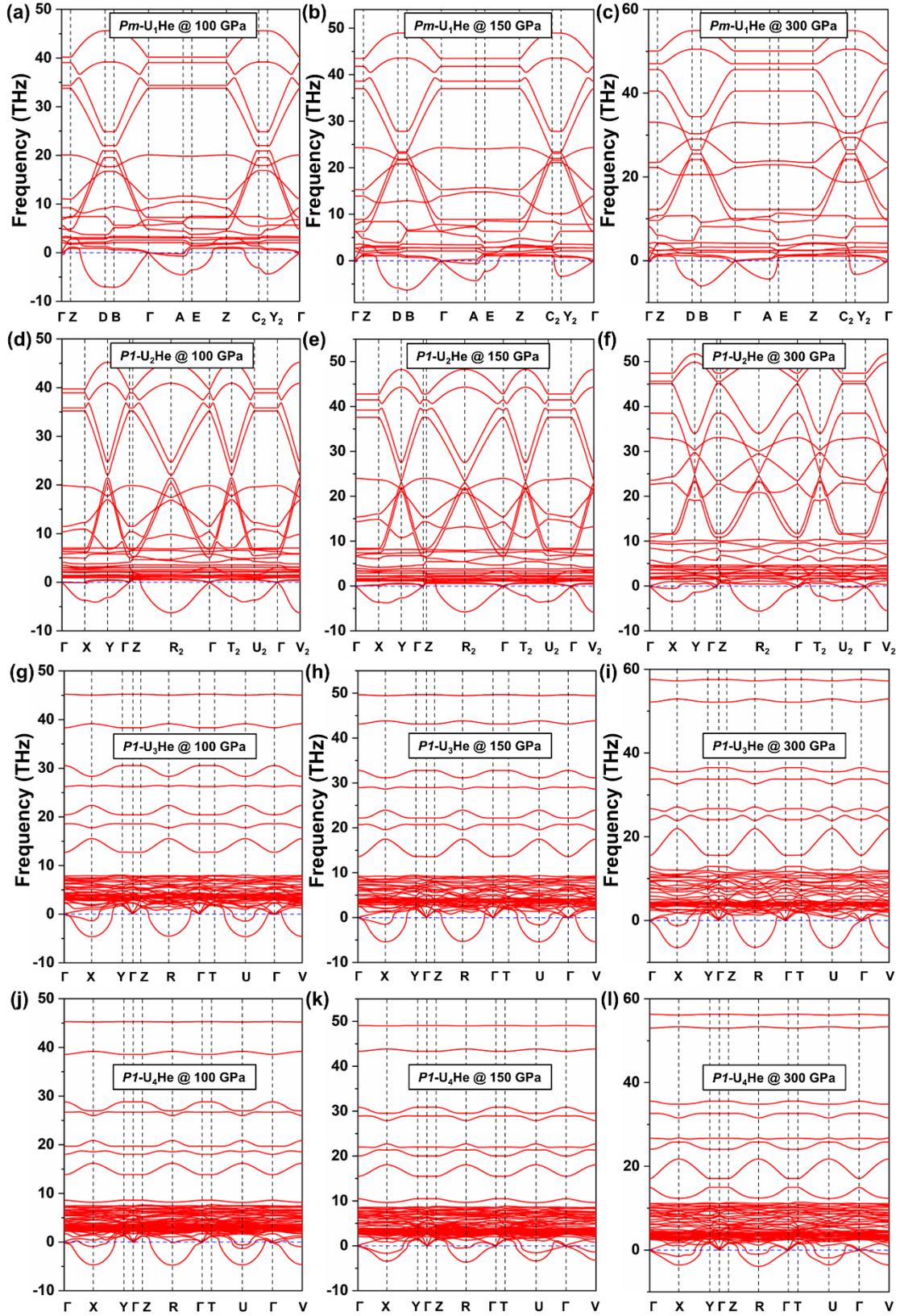

**Fig. S1.** The phonon dispersion spectra at 100, 150, and 300 GPa of (a)-(c) *Pm*-UHe, (d)-(f) *P1*-U₂He, (g)-(i) *P1*-U₃He and (j)-(l) *P1*-U₄He, respectively.



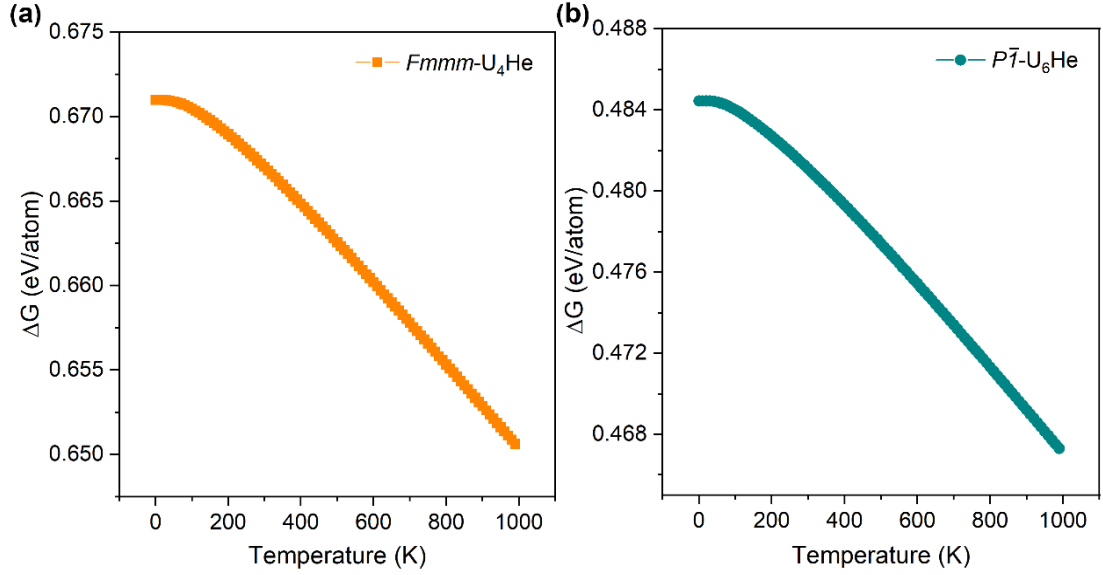

**Fig. S2.** Gibbs free energy of (a) *Fmmm*-U₄He and (b) $P\bar{1}$-U₆He within quasi-harmonic approximation (QHA) at 150 GPa.

Under the quasi-harmonic approximation (QHA), [1] the Gibbs free energy of *Fmmm*-U₄He can be expressed as:

$$\Delta G = \frac{G(\mathrm{U_n He}) - nG(\mathrm{U}) - G(\mathrm{He})}{n+1} \tag{1}$$

where *G(U)* and *G (He)* represented the Gibbs free energy of the most stable structure of element uranium (*α*-U) and helium (*hcp*-He), respectively.

**Table S2.** Bader charge (in e) of *Fmmm*-U$_4$He and $P\bar{1}$-U$_6$He at 150 GPa.

|  | *Fmmm*-U$_4$He | | | | |
|---|---|---|---|---|---|
|  | U1 | U2 | U3 | U4 | U5 |
| Charge | +0.023 | +0.023 | +0.029 | +0.029 | -0.253 |
|  | U6 | U7 | U8 | He1 | He2 |
| Charge | -0.253 | +0.083 | +0.083 | +0.118 | +0.118 |
|  | $P\bar{1}$-U$_6$He | | | | |
|  | U1 | U2 | U3 | U4 | U5 |
| Charge | -0.127 | +0.062 | +0.062 | -0.008 | -0.008 |
|  | U6 | He1 |  |  |  |
| Charge | -0.127 | +0.147 |  |  |  |

**Table S3.** The calculated independent elastic constants $C_{ij}$, bulk modulus $B$, shear modulus $G$, Young's modulus $E$ (in GPa), Poisson's ratio $v$, and longitudinal $v_l$, transverse $v_t$, average sound velocities $v_m$ (in m/s), Debye temperature $\theta_D$ (in K), and $B/G$ of $\alpha$-U under 0 and 100 GPa, as well as experimental data and other theoretical calculations.

| Pressure (GPa) | $C_{11}$ | $C_{22}$ | $C_{33}$ | $C_{44}$ | $C_{55}$ | $C_{66}$ | $C_{12}$ | $C_{13}$ | $C_{23}$ |
|---|---|---|---|---|---|---|---|---|---|
| 0 (Expt.[2]) | 215 | 199 | 267 | 124 | 73 | 74 | 46 | 22 | 108 |
| 0 (Calc.[3]) | 296 | 216 | 367 | 153 | 129 | 99 | 60 | 29 | 141 |
| 0 (Calc.[4]) | 295 | 215 | 347 | 143 | 130 | 102 | 68 | 25 | 149 |
| 0 (Calc.[5]) | 287 | 220 | 352 | 151 | 117 | 101 | 66 | 28 | 152 |
| 0 | 300.0 | 226.2 | 361.5 | 155.0 | 126.2 | 99.2 | 60.1 | 27.3 | 138.9 |
| 100 (Calc.[5]) | 1165 | 770 | 1020 | 355 | 293 | 310 | 224 | 167 | 453 |
| 100 | 1153.1 | 758.5 | 1108.7 | 367.7 | 312.3 | 309.4 | 209.9 | 200.5 | 428.8 |

| Pressure (GPa) | $B$ | $G$ | $E$ | $v$ | $v_l$ | $v_t$ | $v_m$ | $\theta_D$ | $B/G$ |
|---|---|---|---|---|---|---|---|---|---|
| 0 (Expt.[2]) | 115 | 87 | | 0.20 | | | | 251 | 1.322 |
| 0 (Calc.[3]) | 149 | 108 | 261 | 0.21 | | | | 287 | 1.380 |
| 0 (Calc.[4]) | 147 | 108 | 261 | 0.204 | | | | 284 | 1.357 |
| 0 (Calc.[5]) | 148 | 107 | 259 | 0.207 | 3846 | 2338 | 2583 | 283 | 1.383 |
| 0 | 146.6 | 114.1 | 271.8 | 0.191 | 3893.1 | 2406.0 | 2653.6 | 290.8 | 1.285 |
| 100 (Calc.[5]) | 513 | 320 | 795 | 0.242 | 5895 | 3439 | 3815 | 464 | 1.603 |
| 100 | 515.6 | 333.2 | 822.4 | 0.234 | 5959.5 | 3511.3 | 3897.1 | 473.7 | 1.547 |

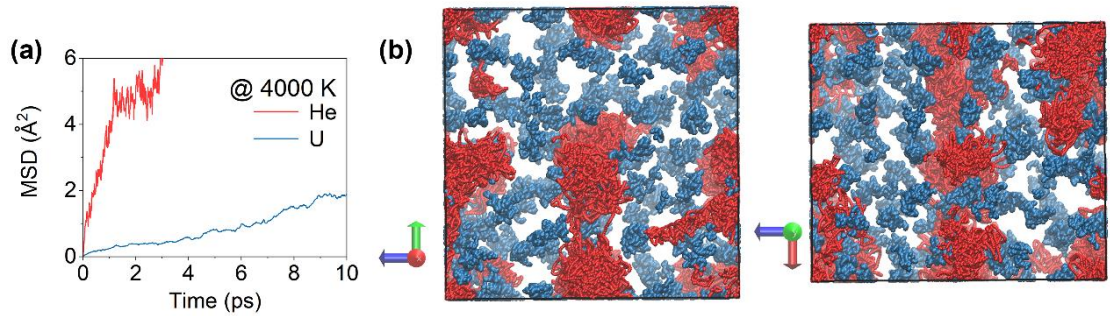

**Fig. S3.** (a) The calculated mean squared displacement (MSD) and (b) the trajectories of U (red) and He (blue) atoms in *Fmmm*-U$_4$He at 4000 K.

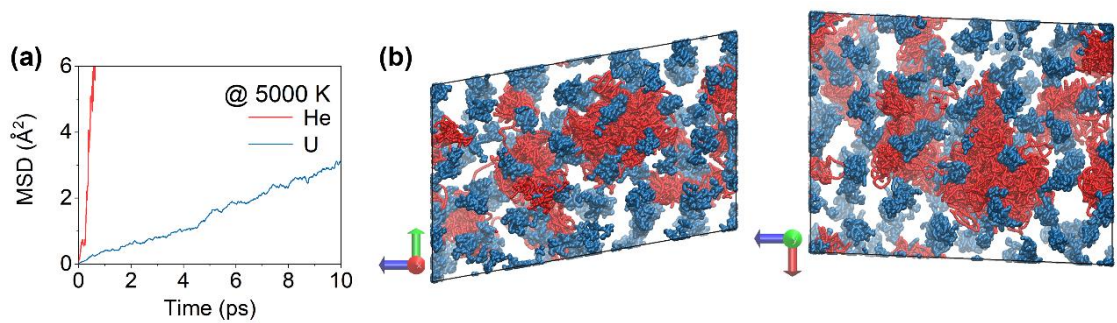

**Fig. S4.** (a) The calculated mean squared displacement (MSD) and (b) the trajectories of U (red) and He (blue) atoms in $P\bar{1}$-U$_6$He at 5000 K.

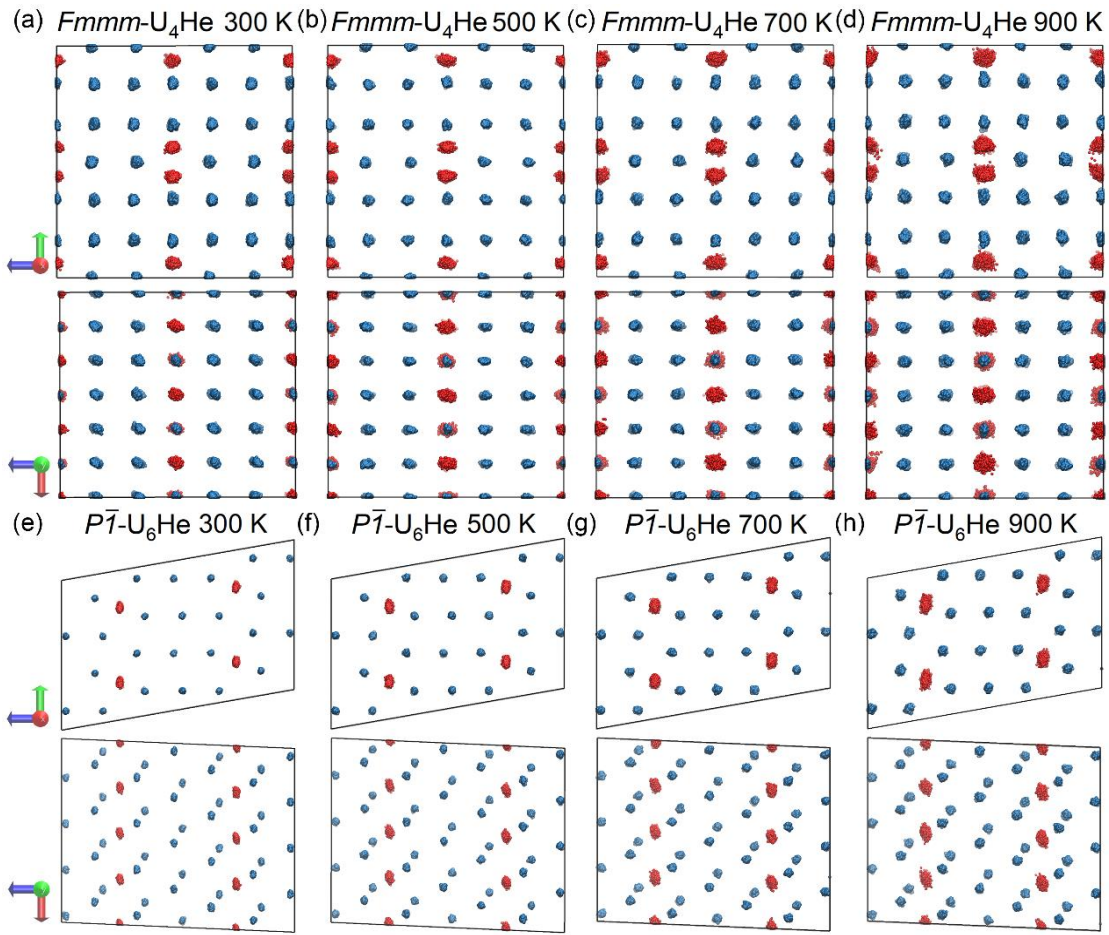

**Fig. S5.** The trajectories of He (red) and U (blue) atoms of (a)-(d) *Fmmm*-U$_4$He and (e)-(h) *P$\bar{1}$*-U$_6$He at 300, 500, 700 and 900 K.

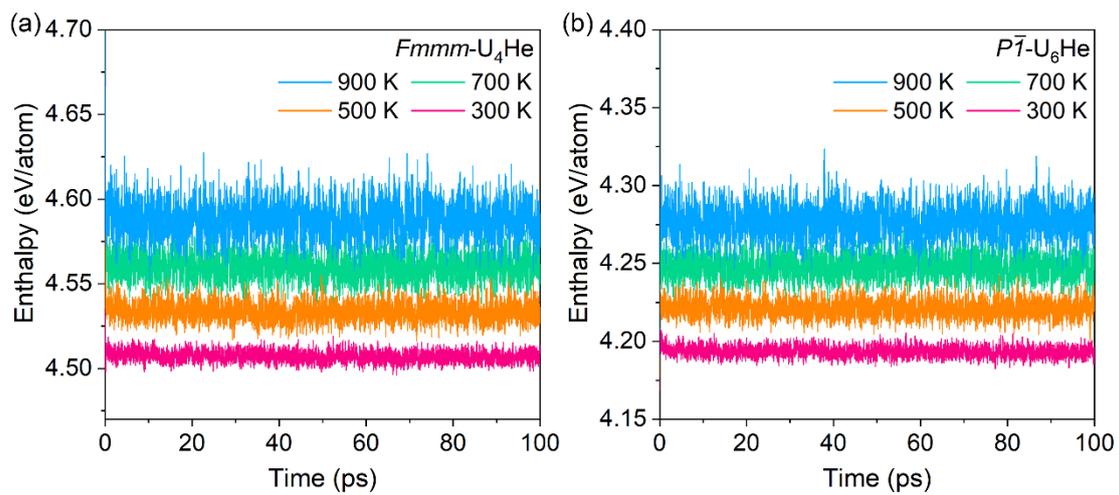

**Fig. S6.** The evolution of enthalpy with time of (a) *Fmmm*-U$_4$He and (b) $P\bar{1}$-U$_6$He at 300, 500, 700 and 900 K.

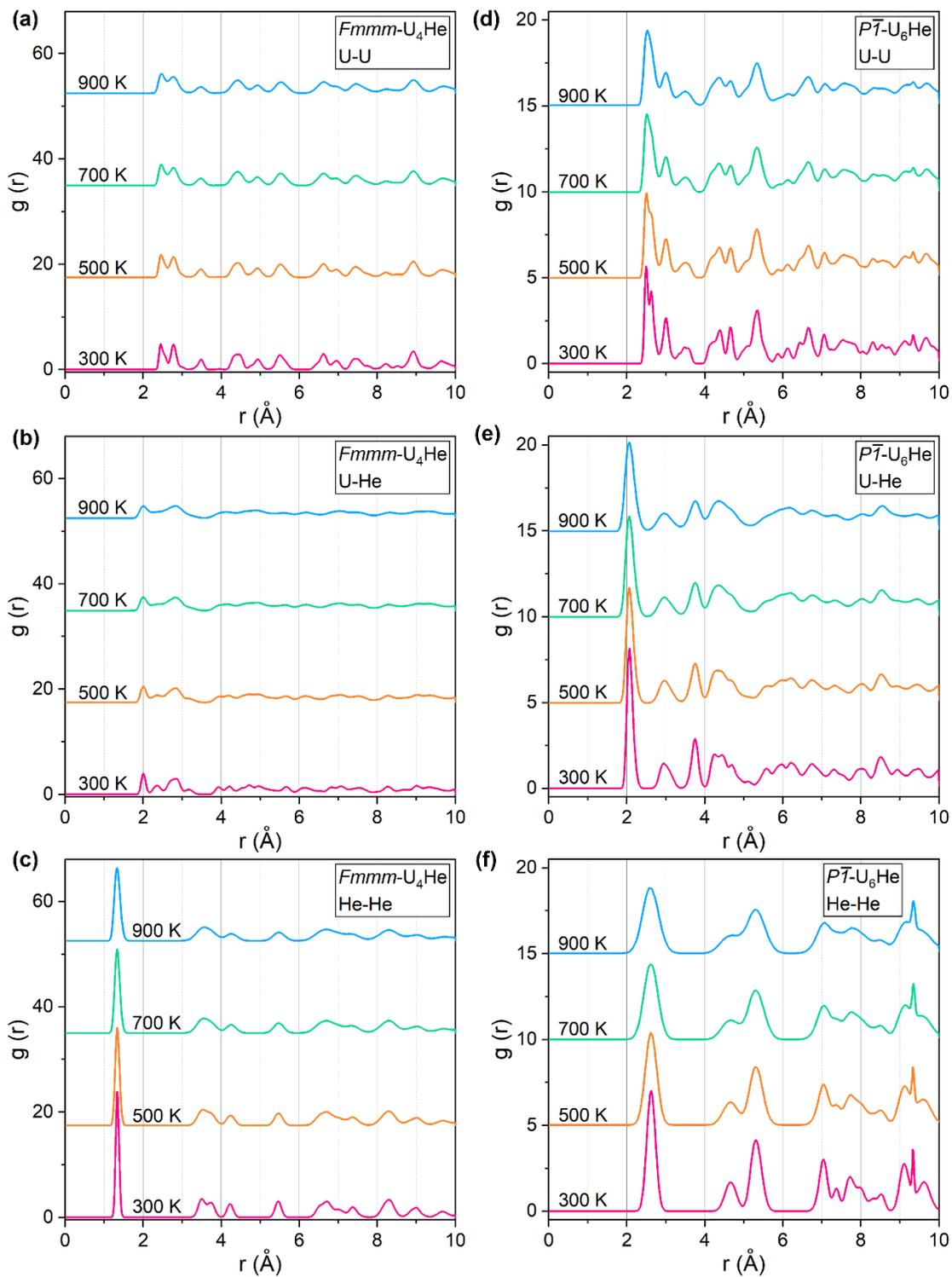

**Fig. S7.** The radial distribution function of (a)-(c) *Fmmm*-U$_4$He and (d)-(f) *P$\bar{1}$*-U$_6$He at select temperature.